\documentstyle[12pt]{article}
\newcommand{\newsection}[1]{
\addtocounter{section}{1}
\setcounter{equation}{0}
\setcounter{subsection}{0}
\addcontentsline{toc}{section}{\protect
\numberline{\arabic{section}}{{\rm #1}}}
\vglue .6cm
\pagebreak[3]
\noindent{\bf  \thesection. #1}\nopagebreak[4]\par\vskip .3cm}
\newcommand{\newsubsection}[1]{
\addtocounter{subsection}{1}
\addcontentsline{toc}{subsection}{\protect
\numberline{\arabic{section}.\arabic{subsection}}{#1}}
\vglue .4cm
\pagebreak[3]
\noindent{\it \thesubsection. #1}\nopagebreak[4]\par\vskip .3cm}

\renewcommand{\theequation}{\thesection.\arabic{equation}}

\newcommand{\ben}{\begin{enumerate}}
\newcommand{\een}{\end{enumerate}}
\newlength{\extraspace}
\newlength{\extraspaces}
\newcounter{dummy}
\newcommand{\bc}{\begin{center}}
\newcommand{\ec}{\end{center}}
\newcommand{\be}{\begin{equation}
\addtolength{\abovedisplayskip}{\extraspaces}
\addtolength{\belowdisplayskip}{\extraspaces}
\addtolength{\abovedisplayshortskip}{\extraspace}
\addtolength{\belowdisplayshortskip}{\extraspace}}
\newcommand{\ee}{\end{equation}}

\newcommand{\ba}{\begin{eqnarray}
\addtolength{\abovedisplayskip}{\extraspaces}
\addtolength{\belowdisplayskip}{\extraspaces}
\addtolength{\abovedisplayshortskip}{\extraspace}
\addtolength{\belowdisplayshortskip}{\extraspace}}
\newcommand{\ea}{\end{eqnarray}}

\newcommand{\ban}{\begin{eqnarray*}
\addtolength{\abovedisplayskip}{\extraspaces}
\addtolength{\belowdisplayskip}{\extraspaces}
\addtolength{\abovedisplayshortskip}{\extraspace}
\addtolength{\belowdisplayshortskip}{\extraspace}}
\newcommand{\ean}{\end{eqnarray*}}
\newcommand{\baa}{                         
\addtocounter{equation}{1}
\setcounter{dummy}{\value{equation}}
\setcounter{equation}{0}
\renewcommand{\theequation}{\thesection.\arabic{dummy}\alph{equation}}
\begin{eqnarray}
\addtolength{\abovedisplayskip}{\extraspaces}
\addtolength{\belowdisplayskip}{\extraspaces}
\addtolength{\abovedisplayshortskip}{\extraspace}
\addtolength{\belowdisplayshortskip}{\extraspace}}
\newcommand{\eaa}{                                       
\end{eqnarray}
\setcounter{equation}{\value{dummy}}
\renewcommand{\theequation}{\thesection.\arabic{equation}}}

\input epsf

\newcounter{fignum}
\newcounter{tabel}

\newcounter{tabnum}
\newcounter{xxx}
\newcommand{\bl}{\begin{list}{({\it\roman{xxx}})}{\usecounter{xxx}}}
\newcommand{\el}{\end{list}}
\renewcommand{\d}{{{\partial}}}

\newcommand{\ppt}[1]{{\partial \over \partial t}}            
\newcommand{\ppx}[1]{{\partial \over \partial x}}            
\newcommand{\pqt}[1]{{\partial^2 \over \partial t^2}}            
\newcommand{\pqx}[1]{{\partial^2  \over \partial x^2}}            
\newcommand{\cN}{{\cal N}}

\def\a{\alpha}

\def\k{\kappa}

\def\<{\langle}
\def\>{\rangle}

\newfont{\gothic}{eufm10 scaled\magstep1}

\renewcommand{\hat}{\widehat}

\begin{document}
\begin{titlepage}
\begin{flushleft}
\today
\end{flushleft}
\begin{flushright}
\vbox{UCU-SCI-02-01}
\end{flushright}
\vskip 5cm
\begin{center}
{\Large\bf Quamtum 2-player Gambling and Correlated Pay-off}
\vskip 1cm
\mbox{F.M.C.\ Witte}\\
{
\it
Department of Sciences, 
University College Utrecht\\
P.O. Box 80145, 3508 TC, Utrecht\\
Netherlands}
\end{center}
\vskip 1.5cm
\begin{abstract}
In recent years methods have been proposed to extend classical game theory 
into the quantum domain. In a previous publication the nature of several
non-commutative games was briefly analyzed. 

Here we give an analysis of the simplest non-commutative 
quantum game, which is a gambling game much like simple heads or tails. The
quantum game displays strategies which, though non direct-product strategies,
allow for correlations between the players pay-off.
\end{abstract}
\end{titlepage}
\newpage

\newsection{Introduction}

Classical Game theory is a core subject in many social sciences oriented 
disciplines  \cite{SSCGames}. Its origins were presented early on in the 
book of von Neumann and Morgenstern \cite{Neumann}. A highly valuable 
step forward in the understanding of the dynamics of games was made by 
John Nash \cite{Nash}. He introduced well-defined equilibrium states,
the so-called Nash equilibria, that play an essential role in defining
the solution of any game. It is however known that not all games possess
a unique Nash equilibrium rendering them essentially unsolvable.

In recent years, roughly since 1999 it has been attempted to quantize
classical games \cite{earlyQG}. A classical game can be quantized by 
assuming that the possible strategies a player can choose are elements of a 
Hilbert space. In the traditional view, players apply unitary operations to 
their strategies, called "tactics", in order to come to a final state that is
then measured by an arbiter for the determination of pay-off. 
The physical model that guides such considerations is that of "classical" 
players communicating their strategies to the referee by means of "quantum 
objects". There are strong indications that entangled states play a significant 
role in finding solutions to quantum games \cite{QBS, entangled, ESS}.

In a recent paper \cite{WittePQG} I have adopted the philosophy of \cite{Qzerosum}
to percieve the programme of quantizing games in terms of quantizing players. 
There are indications \cite{Qbrains} that suggest the human mind may be subject 
to quantum-like fluctuations that give rise to quantum-like behaviour. Generically 
quantum players playing a classical game will obtain the same results as classical 
players playing a quantum game. A possible exception could be cases in which the
players are fermionic, i.e. games where a particular strategy can be played by atmost 
one player \cite{Qbargain}.

An important point of criticism raised recently \cite{EnkPike} concerns two questions.
First one should wonder whether the solutions to quantum games that arise through the 
quantization of classical games have anything to say about the actual game. Secondly,
if entangled states are significant in obtaining new solutions one wonders how players
that cannot communicate according to the rules of the game, still end up in an 
entangled state. In \cite{TQG} the attempt is made to formalize the quantization
procedures and to prove quantum versions of the minimax theorem and the existence of
Nash equilibria for quantum games. This proposal is, as usual, based on the game 
proceeding through the exchange of quantum bits between players and the arbiter.
Consequently gamespaces are automatically direct-product spaces of the strategic
space of the individual players. Als argued in \cite{WittePQG}, this is restrictive
and at the same time does not effectvely counter the more fundamental objections put
forward in \cite{EnkPike}. For quantum players there is no obvious need for their 
pay-off operators to commute. With classical players there is, yielding diagonal 
pay-off operators and allowing one to retrieve the classical limit of quantum 
games in the sector of direct-product states \cite{QBS}. The quantumness of the 
game then shows through the efficiency of entangled strategies, or its influence 
on the formation of coalitions among players \cite{coopQG}. The motivation for 
the introduction of non-commuting pay-off was discussed at length in \cite{WittePQG} 
and will not be repeated here. But let me remark that non-commutative pay-off allows 
for a constructive approach toward quantum game theory, very much in spirit to the 
use of symmetry groups in particle physics phenomenology. The guiding principle used
here is a quantum version of van Enk and Pike's statement that {\it games are defined
by their rules}. Quantum gamespaces should be constructed from the pay-off algebra.
Can that be done?

This paper focusses on a particularly simple game that has been develloped as a 
toy-model for non-commuting games. The 2-player non-commutative game, in 
which the two pay-off operators are essentially canonically conjugate, is 
the game equivalent of the quantum harmonic oscillator. This was noted but 
not discussed at any length in \cite{WittePQG}. This paper will fill that 
ommision. In \cite{QChinos} a Chinos gambling game was analysed. After 
suitable quantization this game shows dynamically generated earnings for one of 
the two players. Here we will see that the non-cammutativity of pay-off also
does this trick for a quantum-like Heads or Tails. In section 2 I analyze this 
2-player non-commutative game. I will construct the game from the operators 
representing the pay-off for two players. Their algebra can be deduced from the
requirement that any operator representing either the actions or measurements by the 
arbiter, or the dynamics of the game, can be constructed from them. The simplest
possible choice generates quantum Heads or Tails. I will give a possible way to 
physically play such a game, and I will analyze the structure of the pay-off earned 
by both players in the process of playing the game.

\newsection{2-player Quantum Heads or Tails}
Let me summarise the ingredients for the description of an N-player quantum game. 
I identify the following parts;
\bl
\item {\bf N-Player Game space:}  The state of the N-player game can be represented by
a vector in a suitable Hilbert space.
\item {\bf Pay-off:} There exists a set of N, linear, self-adjoint pay-off operators
on the N-player Hilbert space, one for every player present.
\item {\bf Arbiter:} There exists an arbiter who determines when the moves should be 
played, and determines pay-off. The operators representing these tasks of the arbiter
can be constructed from the pay-off operators only.
\el
A consequence of this setting is that the arbiter, who determines the rules, indeed 
determines the game. As such this is in spirit with game theory.

\newsubsection{A non-commutative 2-Player betting Game}
A physical model for this game has been presented elsewhere and I will not repeat
that discussion \cite{WittePQG}. Basically such a game could occur when the arbiter 
has posession of a collection of instable nuclei of species $A$ that can decay with equal 
probabillity into species $B$ \{heads\} and $C$ \{tails\}. Two players place a bet on
the occurrence of either heads or tails, with a fixed amount of money for every occurence
of a decay. The arbiter does not know the number of rounds played, i.e. the number of 
decays that have occured, untill the moment he measures it and counts the number of 
heads and tails and determines the pay-off for each player. In a Schr\"odinger Cat 
like case, superpositions of states representing a different number of played rounds 
can occur. They turn this game which classically is a probabillistic gambling game 
into a game where correlations between the two players pay-off may occur.

The arbiter is represented by two operators. The operator $\hat{N}$ counts the sequential 
number of the moves played by the players. Obviously it must be a hermitian operator.
Then there is a raising operator $\hat{\a}_{+}$ that raises this sequential 
move-number by one unit. There is no need for it to be hermitian. In order 
for this to work, the two operators satisfy,
\be
[\hat{N},\hat{\a}_{+}] = \hat{\a}_{+} \ .
\ee
Now let,
\be
\hat{\a}_{-} = \hat{\a}_{+}^{\dagger}
\ee
so we can write $\hat{N}$ as
\be
\hat{N} = \hat{\a}_{+} \hat{\a}_{-} \ .
\ee
The operators $\hat{\a}_{+}$ and $\hat{\a}_{-}$ satisfy
\be
[ \hat{\a}_{-},\hat{\a}_{+}] = 1 \ . 
\ee
From this last result it is obvious that these operators are the well known 
ladder-operators in the quantum mechanics of the harmonic oscillator.

The two ladder-operators are supposed to be related to the pay-off operators 
of the players. Assuming that they are {\it linearly} related, given the fact 
that the pay-off operators are hermitian, the two sole possible combinations 
are straightforward,
\be
\hat{\a}_{\pm} = \frac{1}{\sqrt{2}} [ \frac{1}{\k_{1}} \hat{\pi}_{1} 
\pm \frac{\imath }{\k_{2}} \hat{\pi}_{2} ] \ ,
\ee
where $\k_{j}$ are suitable units of pay-off for the corresponding player. 
It is straightforward to show they satisfy
\be
[\hat{\pi}_{1},\hat{\pi}_{2}] = \imath \k_{1} \k_{2} \ .
\ee
The operator $\hat{N}$, by construction, has a discrete eigenvalue spectrum. 
When we start from a game state that satisfies
\be
\hat{\a}_{-} | 0 \rangle  = 0 \ .
\ee
The eigenvalues take positive integer values. Thus we create a tower of states 
that are labelled by a single integer
\be
| n \rangle  = \frac{1}{\sqrt{n!}} \hat{\a}_{+}^{n} | 0 \rangle  \ ,
\ee
and this set of states is a complete orthonormal set. The corresponding eigenvalues 
are
\be
\hat{N} | n \rangle  = n | n \rangle
\ee
It is straightforward and convenient to compute the matrix representations of these
operators in the basis of eigenstates of $\hat{N}$. In terms of these matrices the 
two pay-off operators become
\be
[ \hat{\pi}_{1} ]_{m n} = \k_{1} ( \sqrt{\frac{n+1}{2}} \  \delta_{m  \ (n+1)} \ 
+ \sqrt{\frac{n}{2}} \  \delta_{m  \ (n-1)} ) \ ,
\ee
and
\be
[ \hat{\pi}_{2} ]_{m n} = - \imath \k_{2} ( \sqrt{\frac{n+1}{2}} \  \delta_{m  \ (n+1)} \ 
- \sqrt{\frac{n}{2}} \  \delta_{m  \ (n-1)} ) \ ,
\ee
It is obvious now that the pay-off in any of the eigenstates of the number operator 
vanishes. How can we infer the existence of discernable strategies from such a game space?
I will show next that the correlations give rise to a fuller understanding of the structure 
of this game.

\newsubsection{Correlations in Pay Off}
First of all we consider the mean-squared deviation for each of the two pay-offs. It shows 
the following behaviour,
\be
\langle n | \hat{\pi}_{j}^2|n \rangle = (n + \frac{1}{2}) \k_{j}^2
\ee
which is typical of succesive, but statistically independent, gambles only in the limit of 
large $n$. This suggests that for large numbers of rounds the game starts to approach a 
symmetric heads or tails game.

The correlations give a more interesting piece of information. In defining the correlation 
between the pay-off of the two players it is important to know that the product
of the two is not hermitian. First we define a "pre" correlation as
\be
PC[\hat{\pi}_{1} , \hat{\pi}_{2} ]_{m n} = 
\frac{1}{2} [ \hat{\pi}_{1} \hat{\pi}_{2} + \hat{\pi}_{2}\hat{\pi}_{1}  ]_{m n} ,
\ee
in the basis constructed in the previous section. By the correlation we then mean
\be
\langle n | C[\hat{\pi}_{1} , \hat{\pi}_{2} ]|n \rangle   = \langle n | PC[\hat{\pi}_{1} , 
\hat{\pi}_{2} ]|n \rangle   - \langle n | \hat{\pi}_{1}|n \rangle \langle n | \hat{\pi}_{2}
|n \rangle  \ .
\ee
Diagonalizing the correlation would give us insight into the possible relationship between the
pay-off of the two players. This would thus provide a proper clue on how to interpret this game. 
Note that the correlation is not the expectation value of a linear operator, due to 
the subtraction of the disconnected terms. Now, it is relatively straightforward to show
that the eigenstates of the pre-correlation have vanishing expectation values for the pay-off.
The correlation can be considered a linear operator when we intepret it as being defined in 
the much larger space spanned by the direct-product of $| n \rangle$ states. However, if we 
restrict to games that are played a {\it finite} number of times we can easilly determine 
the eigenstates of the pre-correlation operator. Note that such a truncation would not make sense 
in the usual harmonic oscillator context. This restriction can be implemented in two ways through 
a modification of the commutation relations between the step operators. 

First of all, applying the step-up operator to the state $| \cN \rangle$ 
could be made to yield the $0$ vector. This is not the same as the state $| 0 \rangle$!
Instead of the commutation relations given above, we would now have,
\be
[ \hat{\a}_{-},\hat{\a}_{+}]_{n m} = \delta_{n m} - \cN \delta_{\cN n}\delta_{\cN m} \ ,
\ee
where $\cN$ is the maximum number of rounds to be played. All the pre-correlation eigenstates 
have vanishing pay-off expectation values. A computation of the eigenvalues of the correlation 
operator for different values of the maximum number of rounds is straightforward. The general 
result is that, upon normalization, correlations are either $1$, $0$ or $-1$. 

If the the state $| \cN \rangle$ is mapped onto the state $| 0 \rangle$ the process of stepping
through the rounds becomes periodic. This leads to another modification of the commutation rules,
\be
[ \hat{\a}_{-},\hat{\a}_{+}]_{n m} = \delta_{n m} - \cN \delta_{\cN n}\delta_{\cN m} - 
\delta_{0 n}\delta_{0 m} \ ,
\ee
It therefore generates a game where the move of the $(\cN+1)$th round maps the $\cN$th-round state 
back onto the initial state.  Such a periodic game shows partially the same characteristics as the 
finite game. Again we find the same set of possible correlations between the values of the two 
pay-off operators. What is particularly interesting about this last modification of the commutation 
relations is that the periodicity after $\cN$ turns yields pay off operators that commute in 
$| 0 \rangle$-sector, i.e. in the initial state.

As a last element of analysis I would like to introduce the equivalent of the time-independent 
Schr\"odinger equation. Using it we can write down the wave functions corresponding to 
seperate rounds and in that representation it becomes easier to study the correlation 
eigenstates in infinite games.

\newsubsection{Schr\"odinger Equation for the Quantum Game}
We use the following representation of the operators $\pi_{j}$
\be
\hat{\pi}_{1} = \imath \k_{1} \k_{2} \frac{\d}{\d \pi} \ ,
\ee
and
\be
\hat{\pi}_{1} = \pi \ .
\ee
It is straightforward to verify that these satisfy the commutation relations
given earlier. Using the step-operators expressed in terms of the pay off 
operators we can write the following equation for the number operator,
\be
- \k_{1} \k_{2} \frac{\d^2}{\d \pi^2} \Psi_{n}(\pi) + 
\frac{1}{\k_{1} \k_{2}} \pi^2 \Psi_{n}(\pi) = (2n + 1)\Psi_{n}(\pi) \ .
\ee
Using the dimensionless variable $\xi = \frac{\pi}{\sqrt{\k_{1} \k_{2}}}$,
we get
\be
- \frac{\d^2}{\d \xi^2} \Psi_{n}(\xi) + \xi^2 \Psi_{n}(\xi) = (2n + 1)\Psi_{n}(\xi) \ .
\ee
The solutions are expressible in terms of the wellknown Hermite polynomials $H_{n}(\xi)$
\be
\Psi_{n}(\xi) = C_{n} \exp [- \frac{\xi^2}{2}] H_{n}(\xi) \ ,
\ee
where $C_{n}$ is a normalization constant. The probabillity that a measurement of 
$\xi$, i.e. pay-off for player 2, yields a value in the range $\xi$ to $\xi + d \xi$
is given by
\be
P_{n}(\xi)d \xi = \Psi_{n}(\xi) \dagger{(\Psi_{n}(\xi))} d \xi 
= \exp [- \xi^2] H_{n}(\xi)^2 d \xi \ . 
\ee
It is now interesting to analyze the relationship with simple Head or Tails from this 
perspective.

As I pointed out earlier, the $n$ is intepreted as an integer counting the number
of rounds that have been played. The classical game would start off with zero
pay-off for both players. If we consider the wavefunction for the zeroth round state
we see it is a pure Gaussian. It peaks at zero pay-off, yet it has a certain width.
So if the arbiter would measure the initial pay-off, there is a calculable probabillity
that the players receive pay-off.
As the first round has been played the game is in the $n=1$ state. Classically this
would mean, for Heads or Tails, that the players have a 50 percent chance of having won
or lost a unit of pay off. So the classical players would find themselves in one of 
these two states, $\xi = \pm 1$. In the quantum game we could ask where the 
probabillity-density peaks. In some arbitrary excited state this happens at
\be
\frac{\d}{\d \xi} P_{n}(\xi) = exp [- \xi^2] 2 H_{n}(\xi)[2 n H_{n-1}(\xi) - \xi 
H_{n}(\xi)] = 0 \ ,
\ee
where the properties of the Hermite polynomials have been used to convert a derivative 
of a polynomial into a lower-order one. The zeros of $H_{n}(\xi)$ generate the minima 
of $P_{n}(\xi)$, so it is when the second factor vanishes that we have a maximum value
of the probabillity density. For the first excited state this leads to the simple 
equation
\be
2 - 2 \xi^2 = 0 \ ,
\ee
which has the obvious solutions $\xi = \pm 1$. So the quantum game amplitude indeed
peaks at the classical values! If we would have assumed that starting from a gaussian
initial pay-off distribution there would have been a 50 percent probabillity to jump 
one unit of pay-off either forwards or backwards, the resulting probabillity density 
would have the sum of two gaussians, one centred around each of the values $\xi = 
\pm 1$. In that case we would have had a minimum at $\xi = 0$, as the quantum game
has, but not as deep as in the quantum case. We can conclude that the quantum game
describes a game displaying more fluctuations than just an initial uncertainty in 
pay-off. Is this behaviour also present for the higher excited states, i.e. the 
higher number of rounds? The answer is ambiguous as the number of maxima satisfies 
our expectations, yet their location starts to deviate from the classical values. 
The microscopic structure of the game described here cannot be represented by a 
classical Heads or Tails game starting out from gaussian initial pay-off. 
If we would assume that pay-off is not generated in fixed and discrete units per round, 
but rather their size is a gaussian stochastic variable itself, we expect the probabillity 
densities to be convolutions of gaussians, and hence gaussians.

The correlation eigenstates now are wave functions satisfying the following first-order
differential equation
\be
\imath \k_{1} \k_{2} [\xi \frac{\d}{\d \xi} + 1]\Psi_{c}(\xi) = \lambda_{c} 
\Psi_{c}(\xi) \ .
\ee
The solutions to this equation are straightforward to write down, but not very 
convenient to interpret. The states are not normalisable, just as the "free 
player states" that are the eigenstates of any of the two operators $\hat{\pi}_{j}$,
i.e. planewaves in pay-off space. 
The integrals involving the product of a free-player state and the conjugate of
another are linearly divergent and give rise to a delta-function normalisation. 
The integrals involving the correlation eigenstates are logarithmically divergent. 
The unnormalized correlation eigenvalue spectrum is continuous and ranges from 
$-\infty$ to $+\infty$. It is straightforward to show that this set of states 
is indeed complete. 

\newsection{Discussion and Conclusion}
On the basis of the material presented above it seems to reasonable to draw
the conclusion that the game space constructed on the basis of the player 
pay-off operators allows for neutral, as well as for winning/losing strategies.
The neutral strategy essentially decouples the players into two games of Heads or Tails.
The other strategies, while maintaining a stochastic character, correlate the random earnings
of one player with the earnings of the other in either a win-win or a win-lose strategy. 
The key element of the construction of these strategies is allowing for superpositions 
of states representing different rounds of the game.

Now where do we take it from here? First of all, the argumentation presented
is far from exhaustive. For example, the question of how a player can opt for a 
particular strategy has not been posed, nor answered. More importantly, the game 
analyzed in this paper represents the simplest, and possibly the most trivial 
non-commuting quantum game. Instead of going about identifying interesting 
non-commutative games through trial and error, one would prefer having theoretical 
tools that allow one to isolate non-trivial games before one has actually constructed 
and analysed the game space. This will require work that goes beyond the work presented 
here and that of \cite{WittePQG}. Last but not least, more knowledge has to be unearthed 
concerning the dynamics of games and it would be highly interesting to find out whether 
a path integral quantization of extensive classical games can lead to the gamespaces 
discussed in the current formulation of quantum game theory. Preparatory work on this 
subject is currently underway \cite{Wessel}.

\end{document}